\documentclass[]{aa}
\usepackage[dvips]{epsfig}
\usepackage{amssymb,calc}

\def\gro{\mbox{GRO J} 1655-40}
\def\grs{\mbox{GRS } 1915+105}

\def\irasnord{\mbox{IRAS } 19124+1106}
\def\irassud{\mbox{IRAS } 19132+1035}

\def\ss{\mbox{SS } 433}

\def\cm{\mbox{ cm}}

\def\microns{\mbox{ } \mu \mbox{m}}

\def\kms{\mbox{ km\,s}^{-1}}

\def\kpc{\mbox{ kpc}}

\def\Kkms{\mbox{ K\,km\,s}^{-1}}

\def\mJy{\mbox{ mJy}}
\def\Jy{\mbox{ Jy}}

\def\K{\mbox{ K}}

\def\Wattmetremoinsdeux{\mbox{ W m}^{-2}}

\def\erg{\mbox{ erg}}

\def\ergs{\mbox{ erg\,s}^{-1}}

\def\mJyparbeam{\mbox{ mJy\,beam}^{-1}}

\def\deg{^{\circ}}
\def\degp{{\rlap.}^{\circ}}
\def\amin{^\prime}
\def\aminp{{\rlap.}^{\prime}}
\def\asec{^{\prime \prime}}
\def\asecp{{\rlap.}^{\prime \prime}}

\def\heu{^{h}}

\def\hmin{^{m}}

\def\hsec{^{s}}
\def\hsecp{{\rlap.}^{s}}

\def\douzecoabb{^{12} \mbox{CO}}
\def\treizecoabb{^{13} \mbox{CO}}
\def\hcoabb{\mbox{H}^{13}\mbox{CO}^{+}}

\def\csabb{\mbox{CS}}

\def\douzecodeuxun{^{12} \mbox{CO } (\mbox{J} = 2 - 1)}
\def\treizecodeuxun{^{13} \mbox{CO } (\mbox{J} = 2 - 1)}

\def\hcoplus{\mbox{H}^{13}\mbox{CO}^{+ }(\mbox{J} = 1 - 0)}
\def\siodeuxun{\mbox{SiO } (\mbox{J} = 2 - 1, v = 0)}
\def\siotroisdeux{\mbox{SiO } (\mbox{J} = 3 - 2, v = 0)}
\def\siocinqquatre{\mbox{SiO } (\mbox{J} = 5 - 4, v = 0)}
\def\csdeuxun{\mbox{CS } (\mbox{J} = 2 - 1)}
\def\hdeux{\mbox{H}_{2}}

\def\Ta{\mbox{T}_{A}^{*}}

\def\plusoumoins{\, \pm \,}

\def\hdeuxromain{\mbox{H\,{\sc ii}}}

\def\ltsima{\; \buildrel < \over \sim \;}
\def\simlt{\lower.5ex\hbox{\ltsima}}            
\def\gtsima{\; \buildrel > \over \sim \;}
\def\simgt{\lower.5ex\hbox{\gtsima}}            

\def\M{\mbox{Mirabel}}
\def\R{\mbox{Rodr\'{\i}guez}}

\usepackage{aabib99}
\begin{document}

   \thesaurus{08     
              (08.09.02 GRS 1915+105;  
	       08.09.02 SS 433;        
               09.09.1 IRAS 19124+1106;  
               09.09.1 IRAS 19132+1035;  
               09.10.1;  
               13.25.5)} 

   \title{A search for possible interactions between ejections from
GRS 1915+105 and the surrounding interstellar medium}
\titlerunning{Interactions between GRS 1915+105 and its surroundings}
 
  \author{S.~Chaty\inst{1,2} \and L.F.~Rodr\'{\i}guez\inst{3} 
\and I.F.~Mirabel\inst{2,4} \and T.R.~Geballe\inst{5}
\and Y.~Fuchs\inst{2} \and A. Claret\inst{2}
\and C.J.~Cesarsky\inst{6} \and D.~Cesarsky\inst{7,8}
 }

   \offprints{S. Chaty}
   \mail{S.Chaty@open.ac.uk}

\institute{Department of Physics and Astronomy, The Open University, 
   Walton Hall, Milton Keynes, MK7 6AA, 
United Kingdom
\and
Service d'Astrophysique, DSM/DAPNIA/SAp, CEA/Saclay, 
L'Orme des Merisiers, B\^at. 709, F-91 191 Gif-sur-Yvette, Cedex, France
\and 
Instituto de Astronom\'{\i}a, UNAM, Campus Morelia, Morelia, 
Michoac\'an, 58190 M\'exico
\and 
Instituto de Astronom\'{\i}a y F\'{\i}sica del Espacio C.C. 67, 
Suc. 28. 1428, Buenos Aires, Argentina
\and
Gemini Observatory, 670 N. A'ohoku Place, Hilo, HI 96720, USA
\and
ESO, Karl-Schwarzschild Strasse 2, D-85748 Garching-bei-M\"unchen, Germany
\and
Universit\'e Paris XI, Institut d'Astrophysique Spatiale, B\^at.
121, F-91450 Orsay Cedex, France
\and
Max Plank Institut f\"ur Extraterrestrische Physik, Postfach
1603, D-85740 Garching, Germany
}

   \date{Received $<$date$>$ / Accepted $<$date$>$}

   \maketitle

   \begin{abstract}

We have observed an extended region surrounding the first discovered
galactic superluminal source $\grs$, seeking evidence of interaction
between the relativistic ejecta of that object and the interstellar
medium. We find two radio sources axisymmetrically aligned along the
sub-arcsecond relativistic ejecta of $\grs$ and roughly 17$\amin$ distant
from it, which coincide with the luminous IRAS sources 19124+1106 and
19132+1035. We have observed these sources at centimeter (VLA), millimeter
(IRAM 30m), and infrared (ISO, UKIRT, ESO/MPI 2.2m) wavelengths in both
line and continuum emission. At centimeter wavelengths a non-thermal
jet-like feature aligned along the outflow axis is located adjacent to the
inner edge of the southern source. Strong density enhancements are found
in the millimeter tracers CO and $\hcoabb$ at the positions of both
sources and some of the morphology is reminiscent of shock-like
interactions; however, linewidths are narrow. At infrared wavelengths
strong hydrogen recombination lines and weak lines of molecular hydrogen
are observed at the southern source. We discuss these results as possible
evidence of the sought-after interaction, both in terms of the regions
undergoing ongoing shock-heating and in terms of them being locations of
shock-induced star formation. The evidence for each of these is
inconclusive. Millimeter line mapping of a portion of W 50 where the
relativistic jets of the X-ray binary $\ss$ interact with the interstellar
medium shows roughly similar morphology as $\grs$, suggesting that the
phenomena observed at the IRAS sources may not be unusual for such a long
distance interaction.


   \end{abstract}

%

\section{Introduction}

The hard X-ray transient GRS 1915+105 was discovered in 1992 by the 
all-sky monitor WATCH on GRANAT \cite{castro:1994}. It has been
intensively studied since then in many wavebands,
including the radio, where ejected plasma clouds with
apparently superluminal velocities, a common extragalactic phenomenon,
were found for the first time in our Galaxy \cite{mirabel:1994a}.
Material ejected from  $\grs$ would be expected to interact with
the surrounding interstellar medium, providing an opportunity to study
in detail for the first time such a relativistic interaction
(for a review see Mirabel \& Rodr\'{\i}guez \cite*{mirabel:1999}).

In a program to look for these interactions we have detected two compact
sources of bright radio emission, each coincident with a bright IRAS
source near $\grs$. These sources are located axisymetrically with respect
to $\grs$ and at same position angle as its normal sub-arsec ejections,
suggesting that they could be the zones of interaction between the ejecta
and the interstellar medium as described in section \ref{2axisources}. We
have observed these two IRAS sources at near-infrared (ESO/MPI 2.2m,
UKIRT), mid-infrared (ISO), millimeter (IRAM 30m) and centimeter (VLA)
wavelengths. The observations are detailed in section \ref{observations}.
We discuss in section \ref{discussion} the possibility of a physical
association between the ejections of $\grs$ and the two IRAS sources,
comparing some of the results obtained in the putative interaction zones
with new measurements of the interaction between one of the jets of SS~443
and the interstellar medium. Some of the observations and results
described here have been briefly described in Chaty \cite*{chaty:1998},
Chaty et al. \cite*{chaty:2000a} and $\R$ and $\M$ \cite*{rodriguez:1998} 
(hereafter RM98).

\section{The context: two axisymmetric sources around GRS 1915+105} 
\label{2axisources}

\begin{figure}
\centerline{\psfig{file=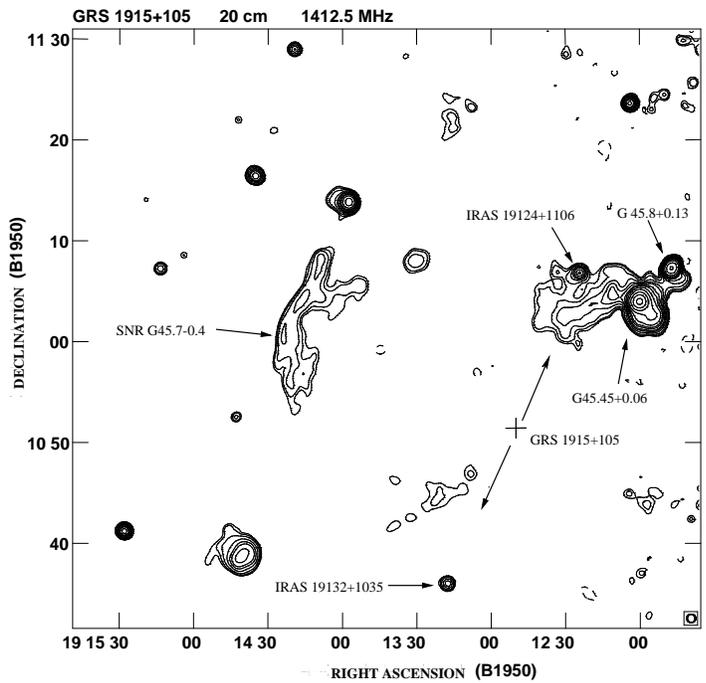,angle=0.,width=9.5cm}}
\caption[]{Map of the surroundings of $\grs$, taken with the VLA-C at
 $\lambda = 20 \cm$. Contour levels are -3, 3, 4, 6, 10, 15, 20, 30, 40, 60, 100, 200, 400 and 800 $\times 3 \mJyparbeam$. The half power contour of the beam is shown in the bottom right corner. The arrows around $\grs$ indicate the position
 angle of the sub-arc relativistic ejecta.
\label{grs_environs}}
\end{figure}

In order to search for interactions involving the energetic and
relativistic ejections of $\grs$, the region surrounding $\grs$ was
observed at radio wavelengths as described by RM98.  This search was
performed at $\lambda = 20$ cm, using the Very Large Array (VLA) of
NRAO\footnotemark\footnotetext{The National Radio Astronomy Observatory is
operated by Associated Universities, Inc., under cooperative agreement
with the USA National Science Foundation}, in its C-configuration, giving
a resolution of $15 \arcsec$. The resulting map is shown in Figure
\ref{grs_environs}.  The region in the map referred to as G45.45+0.06 was
mislabelled G45.46+0.06 in the original papers (e.g., RM98). 
Characteristics of this radio
source are given in Downes et al. \cite*{downes:1980} and in
Feldt et al. \cite*{feldt:1998}.
No evidence of jets or elongated
clouds appears in the figure. However, two small radio continuum sources
positioned nearly axisymmetrically with respect to $\grs$ were
found at angular separations of $17 \arcmin$ each from $\grs$ (RM98). These
small sources are coincident with the bright IRAS sources, 19124+1106 and
19132+1035. Their coordinates are given in Table \ref{position}. The
position angle of the line connecting the northwest source and $\grs$ is
$157 \degp 9$, and their separation is $16 \aminp 6$, the equivalent
values for the southeastern source and $\grs$ are $156 \degp 6$ and $16
\aminp 9$. The position angles are very similar to the position angle of
the sub-arcsec radio-ejections from $\grs$ ($\sim 150 \deg$)
(Mirabel \& Rodr\'{\i}guez \cite*{mirabel:1994a}; 
Fender et al., \cite*{fender:1998a}). The angle between these
ejections and the line of sight towards $\grs$ is $70 \deg$, with the
southeastern ejection approaching and the northwestern one receding
\cite{mirabel:1994a}.

The distance to $\grs$ is crucial for determining many of the
physical parameters of the object (e.g., mass loss rate, 
velocity of ejecta, energetics).
It was estimated to be $12.5 \pm 1.5 \kpc$ by $\M$ \& $\R$ 
\cite*{mirabel:1994a} and by Chaty et al. \cite*{chaty:1996a}.
Recent VLBA observations are consistent
with this value \cite{dhawan:2000}. 
However, the uncertainty may be larger than $1.5 \kpc$, as the above
lower limit was based
on a distance to G45.45+0.06 of $8.8 \kpc$, whereas Feldt et al. 
\cite*{feldt:1998} have placed the H II region at only $6.6 \kpc$.
Moreover, Fender et al. \cite*{fender:1999},
who derived an upper limit of $11.2 \pm 0.8 \kpc$, consider 
that the source is constrained to lie between 7 and 12 kpc.

If the radio sources are at the same distances from the Sun as
$\grs$, and if we assume that the distance of this source is 
$12.5 \pm 1.5 \kpc$, their angular separations correspond 
to the distance of 60~pc from $\grs$.

\begin{table}
\begin{flushleft}
\begin{tabular}{|c|l|l|} \hline
{\em Source} & {\em J2000.0 coord.} & {\em gal. coord.} \\ \hline


GRS 1915+105 
& $\alpha = 19 \heu 15 \hmin 11 \hsecp 545$
& $l^{II} = 45 \degp 40 $                   \\

& $\delta = 10 \deg 56 \amin 44 \asecp 80$
& $b^{II} = -0 \degp 29 $                   \\ \hline


IRAS 19124+1106 
& $\alpha = 19 \heu 14 \hmin 45 \hsecp 77$
& $l^{II} = 45 \degp 54 $                              \\ 

& $\delta=11 \deg 12 \amin 06 \asecp 4$
& $b^{II} = -0 \degp 007 $  \\ \hline


IRAS 19132+1035        
& $\alpha = 19 \heu 15 \hmin 39 \hsecp 13$
& $l^{II} = 45 \degp 19 $    \\ 

& $\delta = 10 \deg 41 \amin 17 \asecp 1$
& $b^{II} = -0 \degp 44 $    \\

\hline
\end{tabular}
\end{flushleft}
\caption[]{\label{position} {\bf Positions of GRS 1915+105, IRAS 19124+1106 and 
IRAS 19132+1035.}
These coordinates are the positions of peak signal at 20 cm as observed by
the VLA. 
}
\end{table}

{\tiny
\begin{table*}
\begin{flushleft}
\begin{tabular}{|llllllllll|} \hline 
{\em Source}   & $1.25\microns$& $2.2\microns$  
               & $12 \microns$ & $25 \microns$ & $60 \microns$ & $100\microns$
               & $2 \cm$       & $6 \cm$       & $20 \cm$ \\
               & mag           & mag                          
               & \Jy           & \Jy           & \Jy           & \Jy
               & \mJy          & \mJy          & \mJy            \\
\hline 
$\irasnord$    & $17.9\pm0.1$  & $12.1\pm0.1$               
               & $3.9$         & $19.6$        & $260.6$       & $581.5$
               & $114\pm6$ & $130\pm4$ & $114\pm4$ \\
\hline 
$\irassud$     & $17.45\pm0.1$ & $10.7\pm0.1$               
               & $6.9$         & $34.0$        & $277.4$       & $488.8$
               & $52\pm6$      & $63\pm4$      & $60\pm4$      \\
\hline 
$\irassud$ ``jet'' &               &                       
               &               &               &               &           
               & $\leq 1$      & $2$           & $5$           \\
%
\hline
\end{tabular}
\end{flushleft}
\caption[]{\label{flux} {\bf Flux densities of IRAS 19124+1106 and IRAS
19132+1035}. \\
All the flux densities with $\lambda$ from $12 \microns$ to $20$ cm come
from RM98.
We refer to Table \ref{isolog} for the fluxes at mid-infrared wavelengths. \\
}
\end{table*}
}

\section{Observations and results} \label{observations}

\subsection{Centimeter}

High-resolution maps of the two radio sources have been obtained with the
VLA. Details of these observations can be found in RM98. These maps, at
three different wavelengths, 2, 6 and 20 cm, are shown in Figure
\ref{les2lobes_vla}.  The morphology of the northern source resembles that
of a cometary HII region, but it also shows a bow shock-like structure to
the South-East, e.g. towards $\grs$. At the southern lobe a non-thermal jet
extends to the northwest along the line between the source and $\grs$. The
flux densities of this jet are $\lesssim 1$, $2$ and $5 \mJy$ at $2, 6$
and $20 \cm$, respectively, indicating a spectral index of $\alpha = -0.8$
($f_{\nu}\propto \nu^{\alpha}$), therefore noticeably different than that
of the rest of the southern radio lobe, which exhibits thermal emission,
as it can be seen in the Table \ref{flux}.
We also reported in Figure \ref{index} a spectral index map of 
IRAS 19132+1035 made from the 20 and 6-cm
maps.
The southern lobe also shows a sharp edge to the south, which could be
either a bow shock, or the ionization front of an $\hdeuxromain$ region.

Additional radio observations of these sources in the $H92 \alpha$
recombination line and adjacent continuum at $3.6 \cm$ are detailed in
RM98. The line strengths, profiles, and radial velocities are typical of
$\hdeuxromain$ regions at kinematic distances of 
$7.4 \pm 1.4 \kpc$ for the north
lobe and $6.0 \pm 1.4 \kpc$ for the south lobe
(distance errors are estimated from the mean value for deviations
of circular rotation of $12 \kms$ given by Brand \& Blitz \cite*{brand:1993}. 
We note that the distance uncertainties are
large because of the low velocity resolution ($\pm 12 \kms$) of the $H92$
observations (RM98).
These estimates are most consistent with the distance to
G45.45+0.06 given by Feldt et al. \cite*{feldt:1998}.
The radio luminosities of the sources derived using these
distances are similar to those of $\hdeuxromain$ regions powered by O9.5
ZAMS and B0 ZAMS stars (RM98).  
If instead they are at the nominal distance
of $\grs$ hotter individual stars or two stars of the above spectral types
would be required to power each source.
However, the interpretation of these objects as
normal $\hdeuxromain$ regions does not explain the presence of the
non-thermal jet-like structure in $\irassud$.
Furthermore, given the large uncertainty in the distance of $\grs$,
we can not {\it a priori} rule out the possibility that $\grs$ 
is roughly at the kinematic distance of the H II regions.

A search for OH maser emission was performed with the VLA in the vicinity of
GRS 1915+105, at 1720 MHz. Such emission can be a signpost for interaction
between a SNR and molecular gas (as in the case of IC443, e.g.
Denoyer \cite*{denoyer:1979}). 
No detection was obtained. However, this does not imply
necessarily that there is no interaction (Rodriguez, Goss, Mirabel,
private communication).

\begin{figure}
\centerline{\psfig{file=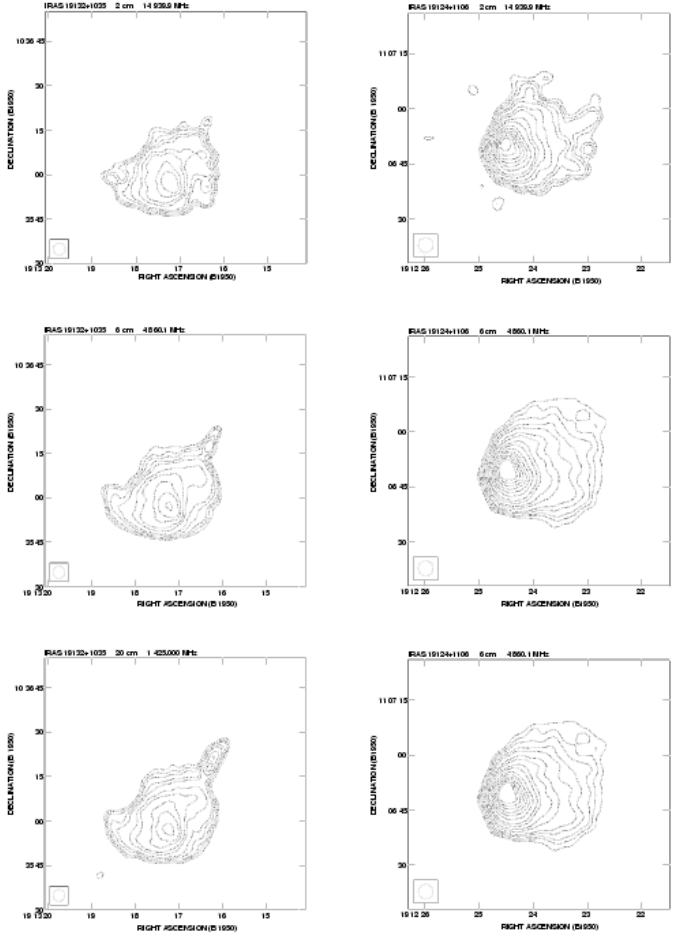,angle=0,width=9.5cm}}
\caption[]{Maps of the two continuum radio sources $\irassud$ (left)
and $\irasnord$ (right), 
acquired with the VLA respectively in the D-configuration 
for the $\lambda = 2 \cm$ map (top), in the C-configuration for the 
$\lambda = 6 \cm$ map (middle) and in the B-configuration for the 
$\lambda = 20 \cm$ map (bottom). 
Contour levels are $-4, 4, 6, 8, 10, 15, 20, 40, 60, 100, 200, 300$ and
$400 \times$ the rms noise of $0.05 \mJyparbeam$.
 The half power 
contour of the beam, with diameter of $4\arcsec$, is shown in each 
bottom left corner. \label{les2lobes_vla}
} 
\end{figure}


        \subsection{Infrared}

                \subsubsection{Near-infrared}

\hspace*{0.5cm}	{\it Broad-band imaging\footnotemark
\footnotetext{Based on observations collected at the
European Southern Observatory, Chile.}} \\

We imaged the IRAS sources in the J ($1.25 \microns$) and K ($2.2
\microns$) bands. The observations were made on 1997, April 5, with IRAC2b
installed on the Max Planck Institute's 2.2~m telescope at the
European Southern Observatory (ESO).
The IRAC2b camera, which contains a Rockwell 256$\times$256 pixel
Hg:Cd:Te NICMOS 3 array detector was mounted at the f/35 infrared adapter
of the telescope. It was used with lens C, providing an image scale of
$0.49 \mbox{ arcsec/pixel}$ and a field of $136 \times 136 \mbox{
arcsec}^{2}$. The typical seeing for these observations was $1.2 \mbox{
arcsec}$. Each final image is the median of 5 frames, each exposed for 2
minutes. After taking each image of the object, an image of adjacent sky
was taken, to allow subtraction of sky emission. The images were processed by
removing bias and dark current, and applying a flat field correction.
These steps were performed using IRAF procedures, in particular the
DAOPHOT package for the photometry in crowded fields.

The images reveal a compact near-infrared counterpart at the position of
$\irasnord$, with $J = 17.9 \pm 0.1$ mag and $K = 12.1 \pm 0.1$ mag, and a
brighter and more extended counterpart of $\irassud$, with $J = 17.45 \pm
0.1$ mag and $K = 10.7 \pm 0.1$ mag. Figure \ref{lobe_sud_j_k} shows the
central portions of the images containing $\irassud$. \\

\begin{figure} \setcounter{figure}{3}
\centerline{\psfig{file=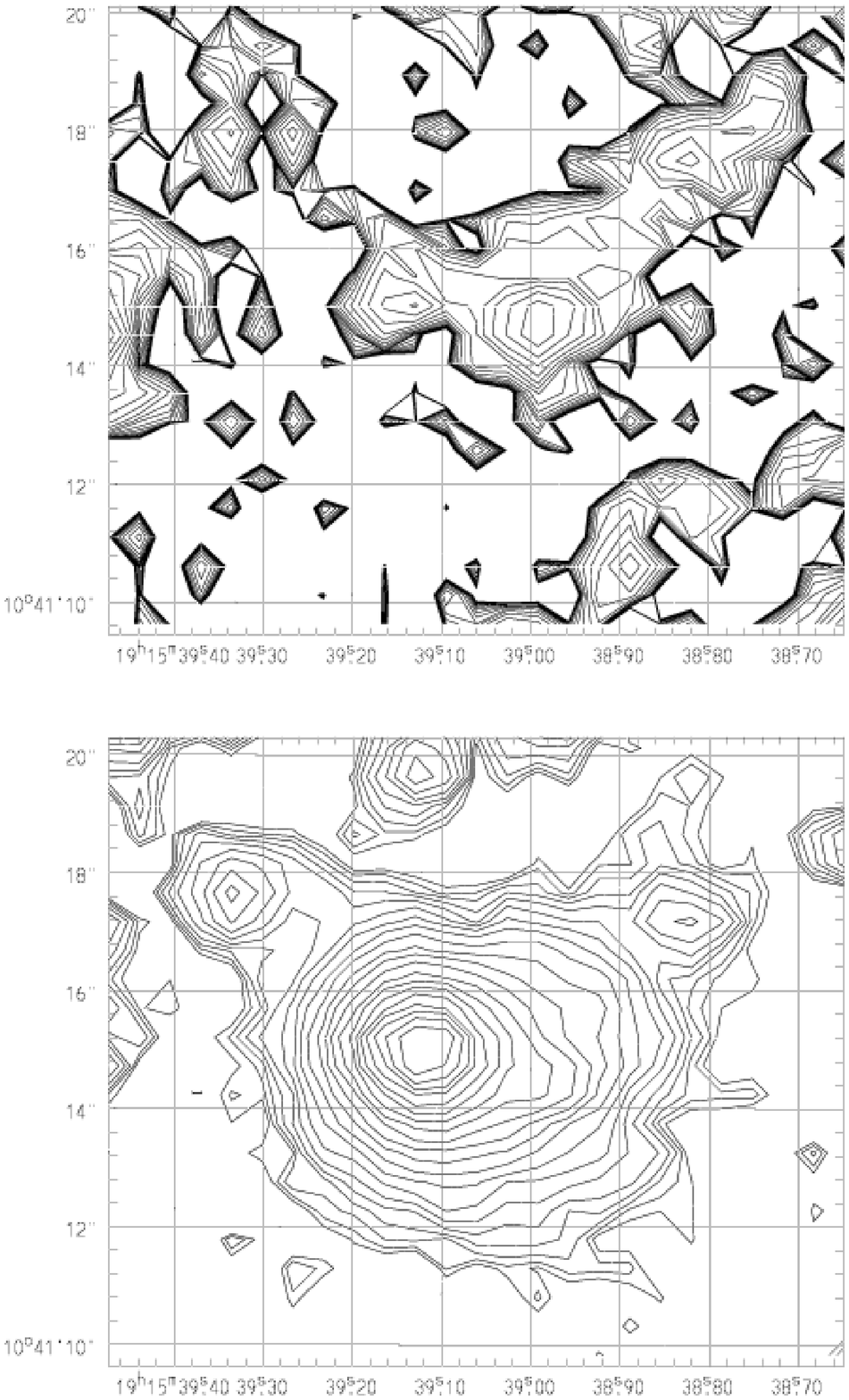,angle=0,width=9.5cm}}
\caption[]{J (top) and K (bottom) band images of the source $\irassud$,
taken with IRAC2b on the 2.2 m of the ESO. The coordinates are for the
J2000 equinox. The ADU flux contour levels are in a logarithmic scale,
each separated by a factor $\sqrt 2$, the first one is equal respectively
to 0.864 for the J band and to 25.501 for the K band.
}\label{lobe_sud_j_k}
\end{figure}

                        {\it Spectroscopy} \\

A low resolution K-band spectrum of $\irassud$ was obtained with the
facility instrument CGS4 on the United Kingdom Infrared 3.8~m Telescope
(UKIRT) on UT 1997 July 13, as part of the UKIRT Service Program. The
75~l/mm grating was used in CGS4 together with a slit of width $1.2 \asec$
to provide a spectral coverage of $0.67 \microns$ at a resolution of
$0.0026 \microns$ ($370 \kms$ at $2.15 \microns$) on the $256 \times 256$
array of InSb detectors. The slit was oriented at a position angle of $148
\deg$ and the telescope was moved to center the peak signal from the
infrared counterpart in one row of the array. Observations were obtained
in the standard stare / nod-along-slit mode. The total integration
time was 8 minutes. A near simultaneous spectrum of the F3V star HR 6987
(T=6700~K, K=4.50 assumed) was obtained at the same airmass; in order to
remove telluric absorption lines (the Brackett $\gamma$ (7-4) absorption line, 
with a
central depth of 0.88 of the continuum at this spectral resolution, was
artificially removed from the spectrum of the comparison star prior to
ratioing). Wavelength calibration was obtained from observations of an
argon arc lamp and is accurate to $0.0005 \microns$.

\begin{figure}
\centerline{\psfig{file=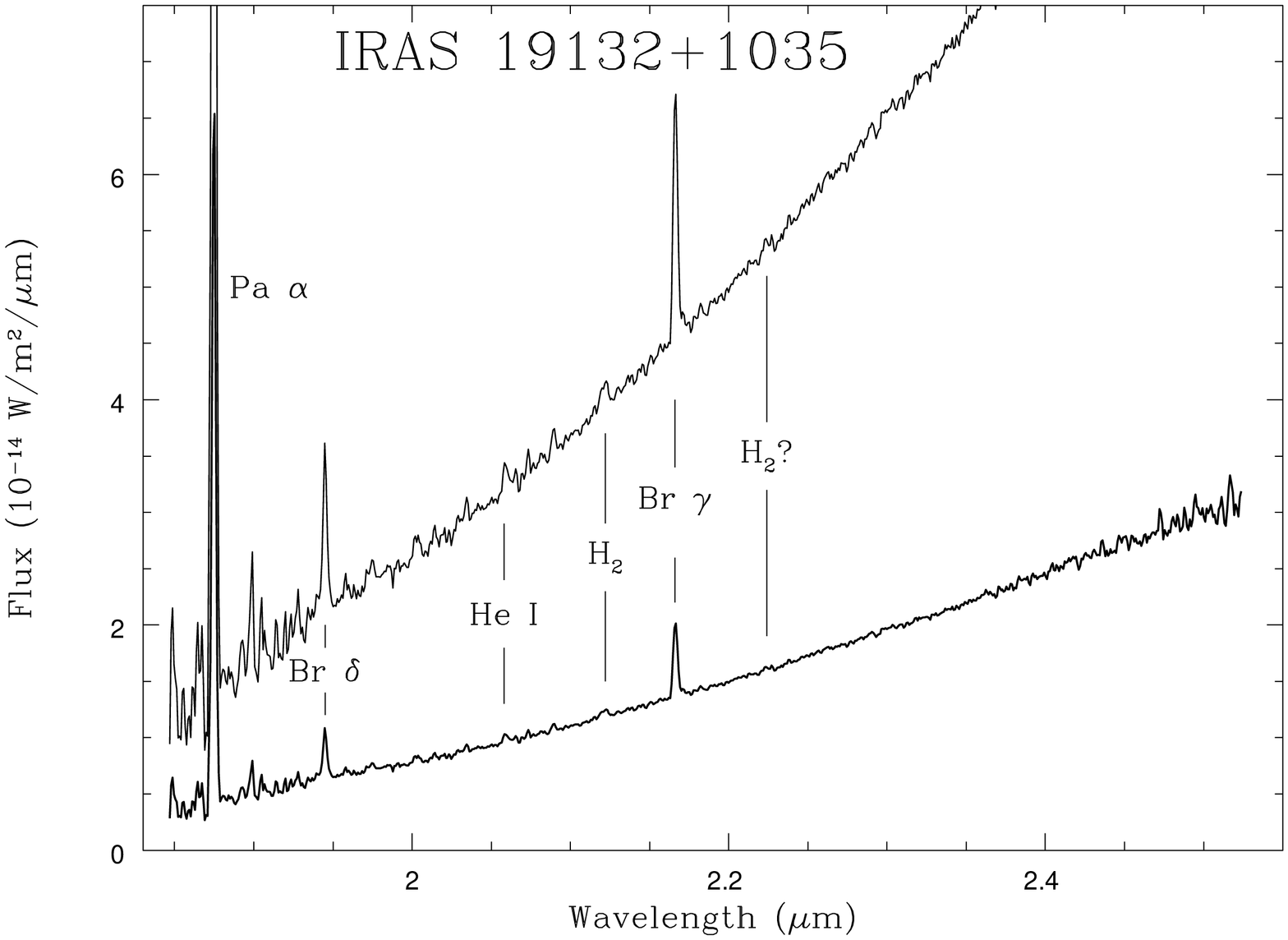,angle=-90,width=10.cm}}
\caption[]{K-band spectrum of $\irassud$ obtained with CGS 4 on UKIRT. 
The upper trace is multiplied by a factor of 3.33 to allow an
easier reading of the weakest lines. Identifications
and positions of detected and marginal lines are indicated.
}\label{lobe_sud_spectre}
\end{figure}

The spectrum of $\irassud$ is shown in Figure \ref{lobe_sud_spectre}. It
is the sum of the spectra of three adjacent rows, covering an area of
$1.2\times3.6$ arcsec. Most of the flux is contained in the central row.
The spectrum exhibits a very red continuum and a number of emission lines.
Most prominent are three recombination lines of atomic hydrogen: a very
strong Paschen $\alpha$ (4-3) line at $1.876 \microns$, and weaker
Brackett $\gamma$ (7-4) and Brackett $\delta$ (8-4) lines, at $2.166
\microns$ and $1.945 \microns$, respectively. The Pa $\alpha$ and Br
$\delta$ lines are uncorrected for the absorption due to the same
line in the comparison star and both it and the Br $\delta$ line occur 
near strong telluric absorption lines of water vapor; thus their strengths
and profiles are subject to systematic errors.

The peak of the Br $\gamma$ emission line occurs at
$+45~\pm~70~\kms$ in the Local Standard of Rest (LSR).  The Full Width at
Half Maximum (FWHM) of the profile is only marginally broader than the
instrumental resolution, but the profile appears to exhibit a weak
high-velocity wing, displaced towards the red. No such wing is present in
the profile of either of the other two atomic hydrogen lines. We note that
lines displaced towards the blue are much more common in the H II regions,
because the front part of an expanding H II region is more easily
detected.

Two fainter emission lines also are visible: the singlet He I $2P-2S$
transition at $2.059 \microns$ and the $H_{2} \, 1-0 \, S(1)$ line at
$2.122 \microns$. The flux in the $\hdeux$ line is $\sim~1~\times~10^{-17}
\Wattmetremoinsdeux$. There is also evidence for the presence of the
$H_{2} \, 2-1 \, S(1)$ line at $2.224 \microns$, at roughly one-fourth the
strength of the 1-0 line, which suggests that most of the excitation of
$H_{2}$ is due to absorbed UV radiation rather than collisions. The 1-0
line appears partly resolved with a deconvolved FWHM of $\sim 500~ \kms$,
but new measurements are required to confirm this. The signal-to-noise
ratio of the 2-1 line is too low to determine if its profile is resolved.

                \subsubsection{Mid-infrared}

Both infrared sources were imaged in six wavebands from 7 to $15 \microns$
with the infrared camera ISOCAM\footnotemark\footnotetext{ISOCAM was
constructed under the scientific direction and the technical expertise of
the Service d'Astrophysique of the CEA/Saclay.  The ISOCAM Consortium is
led by the PI C. Cesarsky.} \cite{cesarsky:1996}, on the Infrared Space
Observatory (ISO)  satellite mission as part of the Guaranteed and Open
Time
Programme. The observations were performed on 1996 April 28 and on 1997
October 20. The log of the observations and the derived fluxes in all
bands are found in Table \ref{isolog}. The images were taken through the
Large Width (LW) filters at angular resolutions given in this Table. 
Data reduction used the CIA package and included subtraction of the dark
current, suppression of the cosmic ray impacts by a multiresolution median
method, correction of the detector transient behaviour, flat field
correction and when necessary a distortion correction (see
Starck et al. \cite*{starck:1999} and references therein).

The observation at $7 \microns$ of the south lobe, shown at the bottom of
Figure\setcounter{figure}{5} \ref{lobes_iso}, 
is particularly striking, because it shows
thermal emission, presumably from heated dust, exactly coincident with the
radio counterpart. One component of this emission is localized, between
the maximum and the jet feature seen at radio wavelengths, on a line
connecting $\grs$ and $\irassud$. The other component is more
extended, and closely resembles the remainder of the radio image with its
bow-shock morphology at the south-east edge. The $15 \microns$ image of
the north lobe is shown at the top of Figure \ref{lobes_iso}. The
morphology is similar to the radio images, resembling a common cometary H
II region, but with two bow shock-like structures facing each other.

\begin{table*}
\begin{flushleft}
\begin{tabular}{c|c|c|c|c} \hline
Source      & Date     & Filter                    & PFOV      & Flux                \\ \hline
$\irassud$  & 28/04/96 & LW2 ($5-8.5 \microns$)    & $3 \asec$ & $13.2 \pm 2.6 \Jy$  \\
$\irassud$  & 28/04/96 & LW7 ($8.7-10.7 \microns$) & $6 \asec$ & $5.9 \pm 1.2 \Jy$   \\
$\irassud$  & 28/04/96 & LW3 ($12-18 \microns$)    & $6 \asec$ & $10.7 \pm 2.2 \Jy$  \\
$\irassud$  & 20/10/97 & LW10 ($8-15 \microns$)    & $1.5 \asec$ & $6.3 \pm 1.2 \Jy$ \\
$\irasnord$ & 20/10/97 & LW9 ($14-16 \microns$)    & $3 \asec$ & $3.3 \pm 0.66 \Jy$   \\
\hline
\end{tabular}
\caption[]{\label{isolog} {\bf Log, characteristics and fluxes
of the sources from ISO observations.} \\
PFOV = Pixel Field Of View. }
\end{flushleft}
\end{table*}

        \subsection{Millimeter}

We used the 30~m radio telescope of the Instituto de Radio Astronom\'{\i}a
Milimetrica (IRAM\footnotemark\footnotetext{IRAM is an European institute
for millimeter astronomy, 
founded by the French Centre National de la Recherche Scientifique (CNRS) 
and the German Max-Planck-Gesellschaft (MPG),
in collaboration with the Spanish Instituto Geografico Nacional
(IGN).}) to observe molecular lines which are good density tracers and are
diagnostics of shock excitation and chemistry. 

The observations, totalling 70 hours, took place during 1997 April 17 --
24, November 29 -- December 3 , and during 1997 December 31 -- 1998
January 5. The observed molecules and transitions, together with their
respective frequencies were: $\douzecodeuxun$ at $\nu = 230.5$ GHz,
$\treizecodeuxun$ at $\nu = 220.4$ GHz, $\hcoplus$ at $\nu = 89.2$ GHz,
$\siodeuxun$ at $\nu = 86.8$ GHz, $\siotroisdeux$ at $\nu = 130.3$ GHz,
$\siocinqquatre$ at $\nu = 217.1$ GHz and $\csdeuxun$ at $\nu = 98.0$
GHz. Details are given in Table \ref{frequences_iram_lobe}.

We used position switching during the observations, with the off-source
position located ($-500 \arcsec$,$-1200 \arcsec$) from the position of
$\grs$.  Tests demonstrated that the off position contained no significant
line emission. The treatment and analysis of the millimetric data were
standard, consisting for each spectrum of a subtraction of a polynomial
baseline of first or second order, in order to remove instrumental
fringes.

\begin{table*}
\begin{flushleft}
\begin{tabular}{lllllllll} 
\hline \noalign{\smallskip}
{\bf Abbr} & {\bf Transition}  & {\bf Frequency}        & {\bf Receiver} & {\bf Banks} & {\bf Beam}\\
\noalign{\smallskip}
\hline \noalign{\smallskip}
$\douzecoabb$  & $\douzecodeuxun$  & $\nu = 230.537\,990$ GHz & 230g2 & b44 b34     & $10.9\asec$ \\
$\treizecoabb$ & $\treizecodeuxun$ & $\nu = 220.398\,686$ GHz & 230g1 & b43 b33 b20 & $11.4\asec$ \\
$\hcoabb$      & $\hcoplus$        & $\nu =  89.188\,523$ GHz & 3mm1  & b42 b32     & $28.2\asec$ \\
$\csabb$   & $\csdeuxun$       & $\nu =  97.980\,968$ GHz & 3mm1  & b41 b31     & $25.7\asec$ \\
SiO 2-1    & $\siodeuxun$      & $\nu =  86.846\,891$ GHz & 3mm2  & b41 b31     & $29\asec$   \\
SiO 3-2    & $\siotroisdeux$   & $\nu = 130.268\,702$ GHz & 2mm   & b42 b32     & $19.4\asec$ \\
SiO 5-4    & $\siocinqquatre$  & $\nu = 217.104\,935$ GHz & 230g1 & b43 b33 b20 & $11.6\asec$ \\
\noalign{\smallskip}
\hline
\end{tabular}
\caption[]{{\bf Frequencies of the transitions and receivers used at IRAM.} 
 The filter bank b3*, divided into sections b31, b32, b33 and b34, provides a low
resolution of 1 MHz, and contains $4 \times 256$ channels, for a total
bandwidth of 1 GHz; the filterbank b20 provides a high resolution of
100 kHz, and contains $2 \times 128$ channels, for a total bandwidth of
25.3 MHz; the autocorrelator b4*, divided into sections b41, b42, b43 and b44
offers resolutions ranging from 10 kHz to 1.25 MHz, a variable number of
channels, and a variable total bandwidth from 20 MHz to 1 GHz.
}\label{frequences_iram_lobe}
\end{flushleft}
\end{table*}

                \subsubsection{$\irassud$}

For each of the $\douzecoabb$, $\treizecoabb$, $\hcoabb$ and SiO 2-1
transitions, we obtained 23 spectra at positions separated by $6 \asec$,
along the line joining $\grs$ to the South lobe and crossing the maximum
of the centimetric continuum shown at the bottom left panel of Figure
\ref{les2lobes_vla}; this maximum was defined as the (0,0) position.
Integration times were 15 minutes per point, except at the (0,0) position
where it was 45 minutes, (-13, 20) where it was 135 minutes, and (-10, 15)
where it was 90 minutes. The two latter locations correspond to the
jet-like feature. In addition we obtained spectra at eight locations
adjacent to the jet. The observations of $\csabb$, SiO 3-2 and SiO 5-4
each consisted of 15 spectra at $6 \asec$ separation along the same line,
with integration times of 15 minutes per position with the exception of
(0,0) and (10,-15) where the times were 30 and 7 minutes, respectively. \\

{\it Results for $\douzecoabb$, $\treizecoabb$, $\hcoabb$ and $\csabb$}
\\

Figure \ref{iras_se_strip2} contains position-velocity plots with an
expanded velocity scale for the observed lines of $\douzecoabb$,
$\treizecoabb$, $\hcoabb$ and $\csabb$. For all transitions the line peak
occurs at an LSR velocity of 67 $\kms$, close to the velocity of $75.7
\kms$ of the H92$\alpha$ recombination line observations as seen by RM98
with much lower velocity resolution.  The intensity distributions of the
millimeter lines are not symmetric, each showing a sharper edge to the
south (positive RA offset), as seen in the VLA images, and a slightly
shifted peak velocity. The $\csabb$ transition, which is a high density
tracer, exhibits a peak shifted to the northwest (more negative RA
offsets), towards the jet-like feature. \\ 

                        {\it SiO} \\

The SiO 2-1 transition was detected only close to the position (-10, 14);
the spectrum at that location is shown in Figure \setcounter{figure}{7}
\ref{iras_se_siodeuxun}.
The line is weak, but its velocity, $67.6 \plusoumoins 0.5\kms$ and width,
$4.4 \plusoumoins 1.2 \kms$ are similar to those seen for much stronger
lines and give us confidence that it is real. The 3-2 transition was not
detected at any individual position, but the average of the spectra
between (0,0) and (-17,25) yields a 3~$\sigma$ detection of a narrow ($1.0
\pm 0.3 \kms$)  line at $67.3~\kms \pm 0.2$, exhibiting an antenna
temperature of $T^{*}_{A} = 0.044$ K and an intensity of $0.05 \pm 0.01$ K
km s$^{-1}$ (Figure \ref{iras_se_siotroisdeux}). The SiO 5-4 transition
was not detected at any individual location or when all of the spectra
were averaged.

\begin{figure}
\centerline{\psfig{file=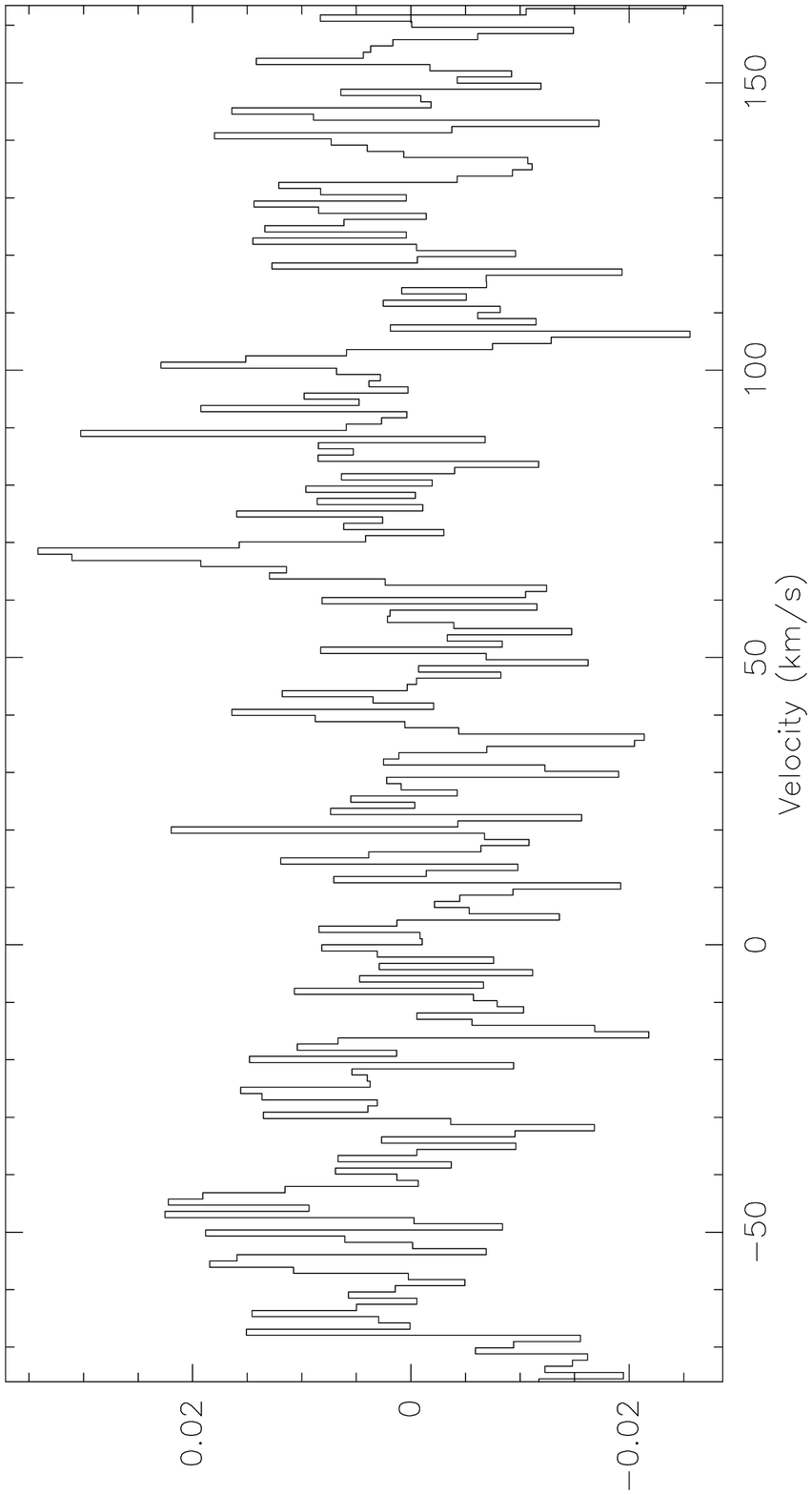,angle=-90,width=10.cm}}
\caption[]{Spectrum of the SiO 2-1 line at IRAS 19132+1035,
obtained at the offset ($\alpha$, $\delta$) = (-$10 \asec$, $+14 \asec$). 
In this and the following spectra, the y-axis is the antenna temperature
in K and the x-axis is the radial velocity with respect to the LSR.}\label{iras_se_siodeuxun}
\end{figure}

\begin{figure}
\centerline{\psfig{file=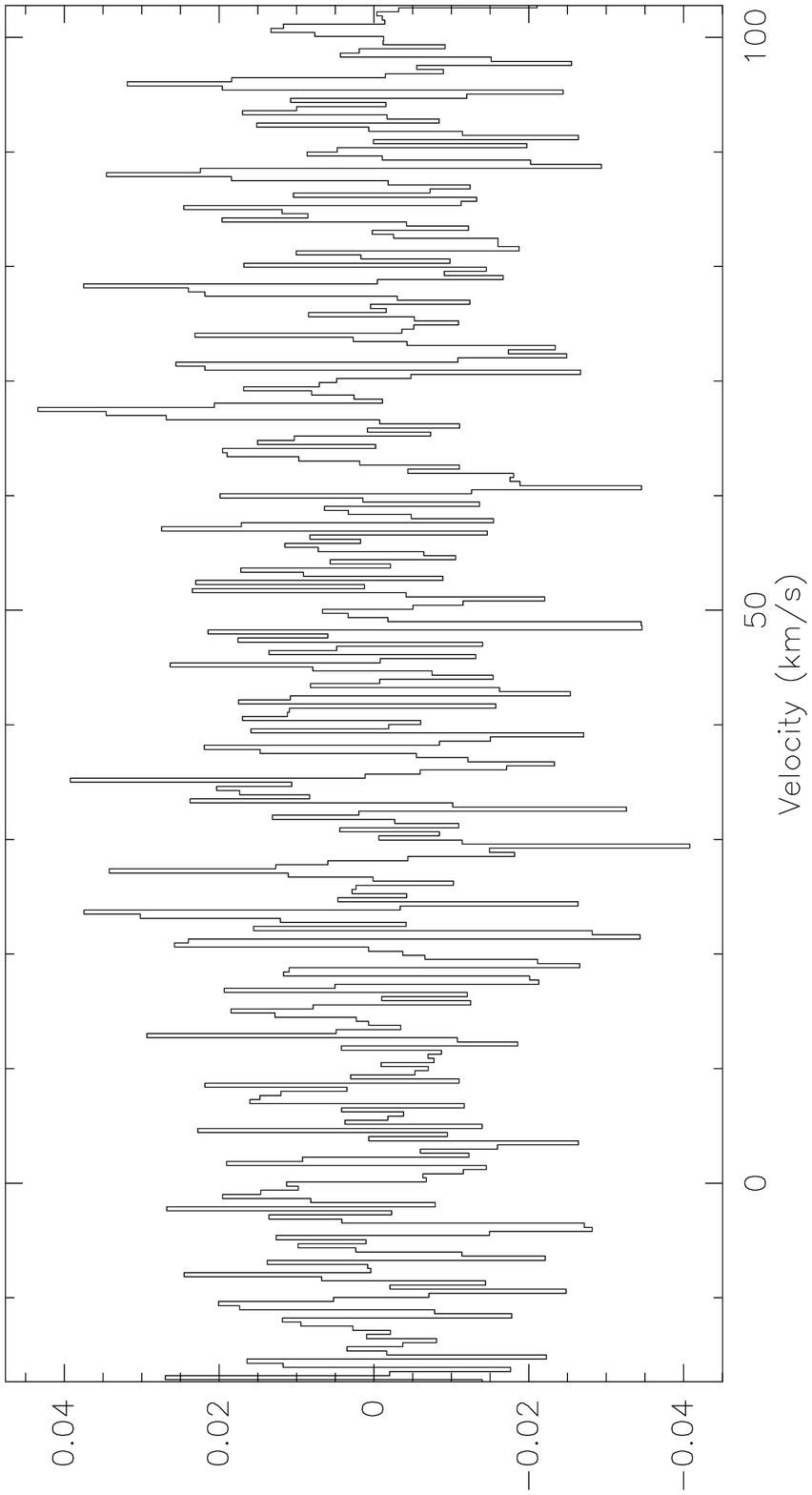,angle=-90,width=10.cm}}
\caption[]{Spectrum of the SiO 3-2 line at IRAS 19132+1035,
averaged over all positions from (0,0) to (-17;25).
}\label{iras_se_siotroisdeux}
\end{figure}

\vspace{5.mm}

                        {\it Discussion} \\

Figure \ref{iras_se_strip2} demonstrates clearly that brightness maxima
for transitions with higher critical densities are displaced increasingly
towards the northwest where the non-thermal jet-like structure is located.
What little SiO emission is detected also originates largely in this
region. The very densest region could be the main interaction zone between
the putative jet and the ambient medium. However, the narrow lines that
are observed are difficult to explain in the above scenario, as an
energetic shock might be expected to lead to line emission over a much
wider range of velocities than is observed. If a shock is present, it
apparently must be of very low velocity.

        \subsubsection{IRAS 19124+1106}

For the $\douzecoabb$, $\treizecoabb$, $\hcoabb$ and SiO 2-1 transitions,
we acquired 24 spectra at positions separated by $6 \asec$, along the line
joining $\grs$ to the northern lobe. As in the case of the southern lobe,
the reference position (0,0) is defined as the position of maximum radio
continuum signal in the northern lobe, as observed by the VLA. The
integration time at each position was 5 minutes, except at the reference
position, where it was 10 minutes. Each of the above transitions was
detected over a wide range of positions. Spectra of the CS 2-1, SiO 3-2
and SiO 5-4 transitions were obtained only at the (0,0) position, each
with an integration time of 15 minutes.  None of the three SiO transitions
were detected at any position at an upper limit of 0.05~K, or when 
all positions were averaged. \\

              {\it Results for $\douzecoabb$, $\treizecoabb$ and $\hcoabb$} \\

Position velocity diagrams for $\douzecoabb$, $\treizecoabb$ and $\hcoabb$
are presented using an expanded velocity scale in Figure
\setcounter{figure}{9} \ref{iras_nw_strip1}.  Two strong velocity components are readily apparent
in all three transitions. For $\douzecoabb$ at the (0,0) position, the
strongest component is centered at $53.95 \plusoumoins 0.03 \kms$, with a
width of $4.05 \pm 0.09 \kms$, an intensity of $12.9 \K$, and an
integrated brightness of $55.5 \pm 0.09 \Kkms$. For $\treizecoabb$, this
component is peaked at $55.10 \pm 0.04 \kms$, and has a width of $3.88 \pm
0.10 \kms$, an intensity of $7.32 \K$, and a integrated brightness of
$30.3 \pm 0.60 \Kkms$. A second velocity component occurs at $59.5 \pm 0.1
\kms$ with a width of $3.9 \pm 0.2 \kms$, an intensity of $3.11 \K$ and an
integrated brightness of $12.8 \pm 0.6 \Kkms$ at the $\douzecoabb$ (0,0)
position. The velocities of these two components are close to that of the
H92$\alpha$ recombination line peak at $57.3 \kms$ (RM98). A third
velocity component, at $\sim 6 \kms$ is visible in the northwestern part
of this cloud for CO and in the southeastern part for $\hcoabb$. Given the
very different velocity the line emission probably originates in a
foreground or background cloud. \\

         {\it Results for $\csabb$} \\

The single spectrum obtained at (0,0), shown in Figure
\ref{iras_nw_cs}, contains a strong component centered at $56.17
\plusoumoins 0.03 \kms$, of width $3.039 \plusoumoins 0.07 \kms$,
intensity $0.937 \K$ and integrated brightness $3.03 \plusoumoins 0.06
\Kkms$. Adjacent to it is a weaker component at $59.91 \plusoumoins 0.11
\kms$, with width $1.589 \plusoumoins 0.22 \kms$, intensity $0.298 \K$ and
integrated brightness $0.504 \plusoumoins 0.07 \Kkms$.  These velocities
are similar to those detected in other molecules and in H92$\alpha$
(RM98). \\

        {\it Discussion} \\

As already noted by RM98 from the centimeter observations, the morphology
of the northern source is that of a cometary H II region.  The millimeter wave
spectroscopy, which reveals two velocity components, is consistent with
this morphology.

\begin{figure}
\setcounter{figure}{10}
\centerline{\psfig{file=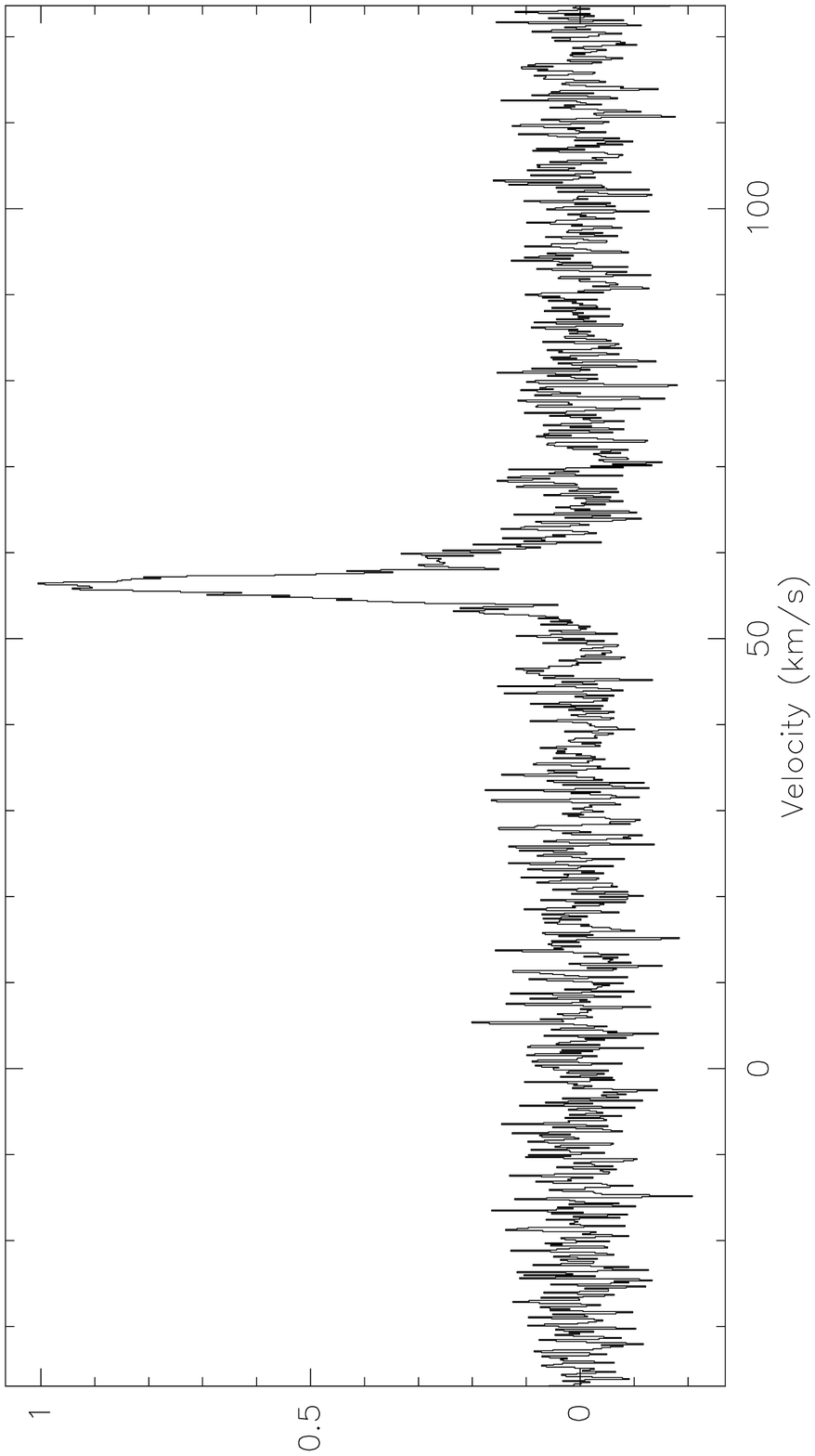,angle=-90,width=10.cm}}
\caption[]{IRAS 19124+1106, $\csabb$, b41, spectrum taken at the position (0,0).
}\label{iras_nw_cs}
\end{figure}



\section{Discussion} \label{discussion}

The new data do not definitively prove or disprove an association between
GRS~1915+105 and either of the IRAS sources.  The strongest lines of
evidence supporting an association are (1) the axisymmetric locations of
the two sources at nearly the same position angle as the recent
sub-arcsecond ejections observed at radio wavelengths, (2) the spatial
coincidence of the non-thermal radio jet with the inner edge of the
southern source, $\irassud$, as well as the orientation of this jet along
the axis, (3) the location of the highest densities in $\irasnord$ and
$\irassud$ on the sides closest to $\grs$ and, in the case of $\irassud$,
close to the non-thermal jet, and (4) the bow shock -- like structure in
the portion of $\irassud$ most distant from $\grs$. Some of these lines of
evidence are associated only with the southern source. Evidence against an
association includes (1) the lack of detected high velocity gas at the
IRAS sources, (2) the luminosities of the IRAS sources, which are
consistent with each of them being powered by one or more hot stars, and (3)
the cometary morphology of the northern source, $\irasnord$, which is
common in star forming regions.

Although the physical alignment of the IRAS sources could be a background
coincidence, the probability of this is low in view of the small number of
IRAS sources, the existence of a jet at $\irassud$, and the locations of
these sources within the Galaxy.  It remains possible that the jet at
$\irassud$ is an extragalactic background source. This should be tested by
radio observations at higher resolution. Because an association between
$\grs$ and these objects may be real we explore in the following the
potential characteristics of interactions of ejecta from $\grs$ with the
interstellar gas and compare these examples with observations of another
somewhat similar source.

\subsection {Production of a large scale cavity}

The angular separation between each IRAS sources
and $\grs$ is $17 \amin$. The measured proper motions 
of the most recent ejections from $\grs$ are 
respectively for the approaching and receding ejecta 
in the interval $\mu_{app} = [17.6,23.6] \pm 0.5 $ mas/d and 
$\mu_{rec} = [9.0,10.0] \pm 0.5 $ mas/d 
($\M$ \& $\R$ \cite*{mirabel:1994a}, Fender et al. \cite*{fender:1999}).
This implies typical travel times to reach the lobes
of 140 years for the approaching ejecta
and of 300 years for the receding ejecta.
It is therefore interesting to note that the impact due to the
South lobe with the southern IRAS source 
would appear to occur much earlier 
than that of the North lobe with the
northern IRAS source.
Evidence for precession of the
jets during the last five years is marginal
\cite{rodriguez:1999a}, with the maximum angle no more than 10 degrees.
Thus a close alignment of currently observed jets close to $\grs$ with
those from long past outbursts that may be responsible for the IRAS
sources may not be a coincidence.

It is perhaps surprising that the separations from $\grs$ of the two
possible interaction zones are identical to better than two percent. These
are determined not only by properties of the outflow from $\grs$, but also
by the density and uniformity of the interstellar medium. In order for
ejected matter to travel up to 60~pc, a large scale cavity must have been
created around $\grs$. Over time this cavity could tend to extend an equal
distance in each direction. The probable existence of a cavity around
$\grs$ already has been pointed out by $\M$ and $\R$ \cite*{mirabel:1996b}. 
The plasma
clouds ejected during the last few years exhibit ballistic motions (i.e.,
with no evidence of deceleration), indicating that the density in the
interstellar medium through which they have passed is less than 0.05
protons cm$^{-3}$ (Mirabel \& Rodriguez, 1996). These clouds have been
observed until their emission faded out or became unobservable with the
available VLA configuration, typically at distances of one arcsecond from
the central source.

Although direct evidence for a cavity is limited to a region very close to
$\grs$, it is likely that the cavity extends well beyond one arcsecond
from it. Apart from the two IRAS sources, no phenomena that might be
associated with deceleration of recent ejecta have been observed from
1$\asec$ angular separation from $\grs$ to 17$\amin$. In order to produce
a cavity with a linear dimension of many tens of parsecs, even a
relativistic source must be active for a much longer period of time than
$\grs$ has been observed. From EXOSAT measurements \cite{reynolds:1999}  
$\grs$ was active in 1985 but much fainter than it has been since 1992. 
For the
second discovered superluminal galactic source, $\gro$, 
Kolb et al. \cite*{kolb:1997}
suggested that the activity of the source is caused by the companion
crossing the Hertzsprung gap, a transient phase lasting $\sim$1~Myr. The
companion of $\grs$ \cite{marti:2000} is believed to be more massive than
the companion of $\gro$ \cite{shahbaz:1999}, and thus the duration of
the active phase would be much shorter, although still sufficiently long,
especially since $\grs$ has been much more active than $\gro$.

\subsection{Interaction between ejecta and ambient material}

Considerable heating of the gas must occur at the ends of the cavities,
where the ejecta, whether or not partially decelerated en route, impact
ambient material. One would expect that much of heating would occur via
shock excitation. For the case of $\grs$ the speed and density of the
incident material and the relative amounts of atomic and molecular gas
(both in the incident and impacted material) are not known, and thus the
spectrum of the shock-heated gas is uncertain.  The southern IRAS source
shows strong infrared lines of atomic hydrogen and weak infrared line
emission from molecular hydrogen, very unlike the emission observed in
molecular clouds impacted by protostellar winds having speeds of up to
100 $\kms$, in which lines of molecular hydrogen dominate and hydrogen
recombination lines are nearly non-existent (e.g. Geballe \& Garden
\cite*{geballe:1987}).

We note that if the non-thermal feature near the southern lobe is an
element of the interaction involving the ejecta of $\grs$, 
the collimation of the jet is $f_{coll} =
\frac{length_{jet}}{width_{jet}} = \frac{15'}{7\asec} \gtrsim 100$, ten
times larger than the most highly collimated Herbig-Haro objects (see e.g.
Bachiller \cite*{bachiller:1996}).
This implies a jet opening angle of $\lesssim 0.5\deg$
which is consistent with the limit of $< 8 \deg$ 
derived by Fender et al. \cite*{fender:1999} based on observations
of the core, but is more highly constraining.
This jet opening angle seems to be in agreement with the jet full
opening angle as function of distance from the core 
for M87 \cite{junor:1999}, 
if we take into account the scaling factor between galactic and extragalactic
black holes.
Therefore, as for extragalactic jets, it would require in the galactic case
ongoing confinement
of the jet at large distances from the source, ruling
out free expansion of a relativistic gas. \\

	{\it Induced star formation?} \\

The similarity of the luminosities of the IRAS sources to those of compact
H~II regions containing one or more massive stars, as well as the spectrum
of $\irassud$, which shows strong hydrogen recombination line emission,
suggest the possibility that ejecta from $\grs$ have induced massive star
formation at the locations of the IRAS sources, via compression of the
interstellar gas. In this scenario the non-thermal radio jet observed at
$\irassud$ might be interpreted as a Herbig-Haro-like feature, the result
of a protostellar wind breaking out of the natal cloud. However, given the
length of time for star formation to proceed to this phase, such an
explanation would require activity from $\grs$ at a much earlier time than
mentioned above. Moreover, there appears to be no reason for star
formation at $\irassud$ to produce a close coincidence between the
position angle of the radio jet at $\irassud$ and that of the line
connecting $\grs$ and $\irassud$, as is observed. Finally, the radio
emission from Herbig-Haro jets usually is thermal (free-free;
$\R$ \cite*{rodriguez:1999b}).  In view of these arguments, we do not regard
induced star formation to be a likely explanation for the luminosities of
$\irasnord$ and $\irassud$. Dubner et al. (1998) note that in the case of
the relativistic jets from SS~433 there is no evidence of induced star
formation in the impacted gas.

\subsection{Comparison with SS 433}

In view of the inconclusive evidence linking the IRAS sources with
activity originating at $\grs$, it is perhaps useful to examine other
possible examples of such an interaction. Like $\grs$, the famous X-ray
binary SS~433 ejects beams of material at relativistic speeds. This object
is located inside the radio shell/supernova remnant W50, which exhibits
two lateral extensions with dimensions of tens of parsecs
\cite{dubner:1998}. The morphology can be attributed to continuous
ejection of magnetic field and high-energy particles from the central
source. Sub-arcsecond jets are present at SS~433, with 5 orders of
magnitude difference between their extents and those of the far lobes. The
power injected into these jets is $\sim 10^{39} \ergs$ according to
Dubner et al. \cite*{dubner:1998}, 
who also estimate that the kinetic energy transferred
to the surroundings of $\ss$ during the last $\sim 2 \times 10^4$ years
amounts to $2 \times 10^{51} \erg$.

Major ejection events in GRS 1915+105 are more sporadic than in SS 433,
and therefore the average kinetic energy injected into the surroundings
over long periods of time may be smaller \cite{mirabel:1999}. The
kinetic energy of the 1994 March 19 event in $\grs$ was $\sim 10^{43}
\erg$ \cite{rodriguez:1999a}. $\grs$ would need to exhibit one ejection
event per hour similar to this one in order to match SS~443. A more likely
rate is one event per month \cite{rodriguez:1999a}, in which case
$\sim~2~\times~10^7$ years would be required to equal the energy output of
SS~443 in $2 \times 10^4$ years.  However, we note also that the jets
emanating from $\grs$ always are compact close to the source
\cite{dhawan:2000}, which indicates that a continuous injection of energy
from $\grs$ is occuring. In addition to the large events described above,
there are also smaller ones, observed at X-ray, radio and infrared
wavelengths, where the mechanical luminosities are in the range
$10^{37}-10^{39} \ergs$, and where the synchrotron emission is seen up to
the infrared wavelengths (e.g. Mirabel et al. \cite*{mirabel:1998a}, 
Fender \& Pooley \cite*{fender:2000}).
A comparison of the energetics of $\grs$ with SS~433 and a Herbig-Haro
object is reported in Table \ref{compar}.

Despite the differences between $\grs$ and SS~433, each may have created a
cavity of similar linear extent.  In view of the uncertainty as to what
effects might be observable at the putative ends of the cavity produced by
$\grs$, we have observed the western tip of W50 where the ejecta of SS~443
and the shell remnant interact, in some of the same millimeter lines of
$\douzecoabb$, $\treizecoabb$, $\hcoabb$ and SiO as were observed for the
IRAS sources. The observed region in W50 is at the constant declination
$05 \deg 00 \arcmin 00 \arcsec$ and right ascensions between $19 \heu 05
\hmin 27 \hsec$ and $19 \heu 06 \hmin 24 \hsec$ (B1950) (see
Dubner et al. \cite*{dubner:1998}, figures 1a and 1b).  
The off position was 15
arcminutes east and north of the reference point located at $19 \heu 06
\hmin 00 \hsec$; $05 \deg 00 \arcmin 00 \arcsec$. Measurements at W50 were
separated by $24 \asec$, with integration times of 5 minutes each, except
at (0,0) where the integration time was 10 minutes.

The results are shown in Figure \ref{w50_strip} over the relevant velocity
range. Emission from $\douzecoabb$, $\treizecoabb$ and $\hcoabb$ are found
roughly 200 -- 300$\asec$ west of the reference position, at the
westernmost edge of the radio shell (Dubner et al. \cite*{dubner:1998}, 
Figures 1a and
1b). The emission from the high density tracer $\hcoabb$ is the most
compact and probably indicates the location of the strongest interaction.
The line profiles of all three species are asymmetric but narrow. SiO 2 --
1 emission was not detected. These results are very similar to those for
$\irassud$. These similarities between SS~433/W50 and $\grs$/$\irassud$
and $\irasnord$ suggest that high density, low velocity molecular clumps
may not be unusual products of the interaction of the ejecta of a distant
energetic source and the surrounding medium.

\begin{table}
\begin{flushleft}
\begin{tabular}{c|c|c|c} \hline
              & HH              & SS 433          & GRS 1915 \\ \hline
Velocity      & 1 $-$ 500 km/s   & 0.26 c          & 0.92 c   \\
Separation    & 1au $-$ 10pc     & 50 pc           & 60 pc    \\
Range         & $10^{6}$        & $10^{5}$        & $10^{4}-10^{5}$ \\
Collimation   & 2 $-$ 10         & $\geq 12$       & $\sim 100$\-? \\
Kinetic Pow.  & $10^{32} \ergs$ & $10^{39} \ergs$ & $10^{36-39} \ergs$ \\
\hline
\end{tabular}
\caption[]{\label{compar} {\bf Characteristics of jets from
different sources.} HH = Herbig-Haro. The range is the ratio of distance
between the far lobes and the subarcsec jets, the collimation factor is
the ratio between the length of the jet and its width, and the kinetic
power of the ejections is a time average.
}
\end{flushleft}
\end{table}

\section{Conclusions}

We have performed extensive multi-wavelength observations from IR to radio
of two radio/IRAS sources axisymetrically located with respect to $\grs$
and aligned with the position angle of the subarcsec jets. The northern
source has the morphology of a common cometary H II region. The
observations of the southern cloud reveal a collimated non-thermal
structure which may be the far end of a jet of material from $\grs$. Both
sources contain dense clumps of molecular material. Overall the evidence
for these regions being interaction zones seems inconclusive. An
abbreviated study of what may be a very similar interaction between the
ejecta of SS~433 and the surrounding interstellar medium reveals
strikingly similar (and also apparently inconclusive) phenomena, perhaps
indicating that such phenomena might be produced by sources which via
their violent behavior have created large cavities over long periods of
time.

\begin{acknowledgements}

We thank F. Comeron for allowing us to use part of his time at the ESO/MPI
2.2 m telescope of the ESO/La Silla. S.C. thanks the astronomers on duty
at IRAM, particularly R. Moreno and D. Reynaud, for much assistance and
many fruitful discussions during the observations.  He thanks also G.
Pineau des For\^ets and B. Le Floch for helpful discussions. S.C. would
like to acknowledge the Instituto de Astronomia de Morelia, group of the
Universidad Nacional Autonoma de Mexico (UNAM), for being invited to work
in this dynamic institute, and for the fruitful discussions leading to
some of the ideas in this paper. He thanks also the Groupe de Recherche
Accr\'etion-Disque-Jets (GdR ADJ) of the French Centre National de la
Recherche Scientifique (CNRS) altogether with the UNAM for the financial
support of this journey.  We thank R.P. Fender for pointing out to us the
previous mislabelling of the region G 45.45+0.06. 
We also thank the anonymous referee for prompt and useful comments,
which allowed us to improve the manuscript.
S.C. acknowledges
support from grant F/00-180/A from the Leverhulme Trust. I.F.M.
acknowledges support from CONICET/Argentina. The United Kingdom Infrared
Telescope is operated by the Joint Astronomy Centre on behalf of the U.K.
Particle Physics and Astronomy Research Council. The ISOCAM data presented
in this paper were analysed using "CIA", a joint development by the ESA
Astrophysics Division and the ISOCAM Consortium.

\end{acknowledgements}

\bibliographystyle{aabib99}
\bibliography{science}

\onecolumn

\newpage

\twocolumn
\newpage


\begin{figure}[Ht]
\setcounter{figure}{2}
\setlength{\unitlength}{1.0cm}
\vspace*{-0.5cm}
\begin{picture}(8,9.)(-0.5,0.5)
\includegraphics{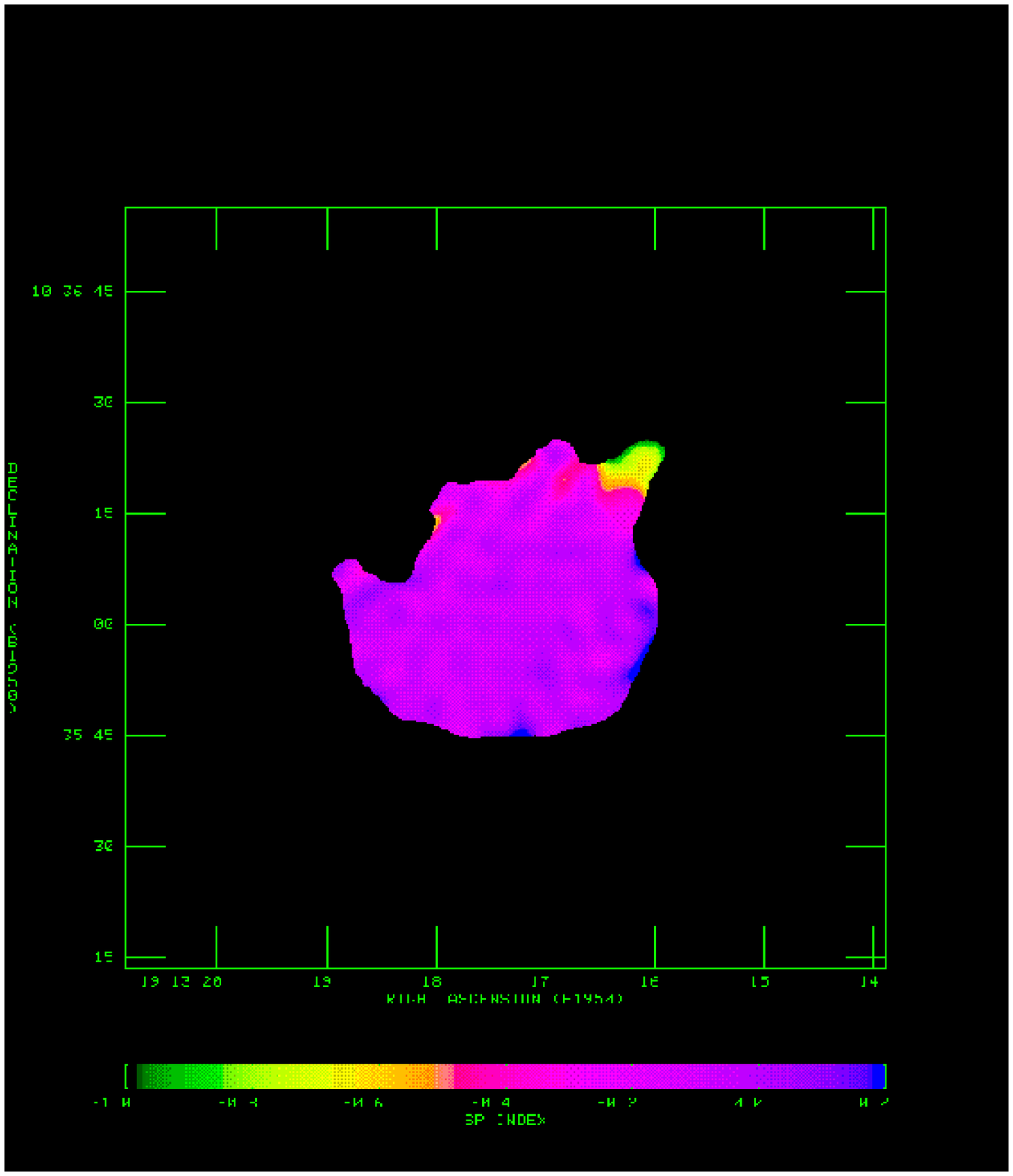}
\end{picture}
\vspace*{-0.5cm}
\caption[]{Spectral index map of IRAS 19132+1035 made from the 20 and 6-cm
maps. The color coding for the spectral index
is given at the bottom of the figure. Note the negative,
non-thermal 
spectral index of the jet feature to the northwest.
\label{index}
} 
\end{figure}

\begin{figure}[Hb]
\setcounter{figure}{5}
\setlength{\unitlength}{1.0cm}
\begin{picture}(8,10.)(0,1.55)
\includegraphics{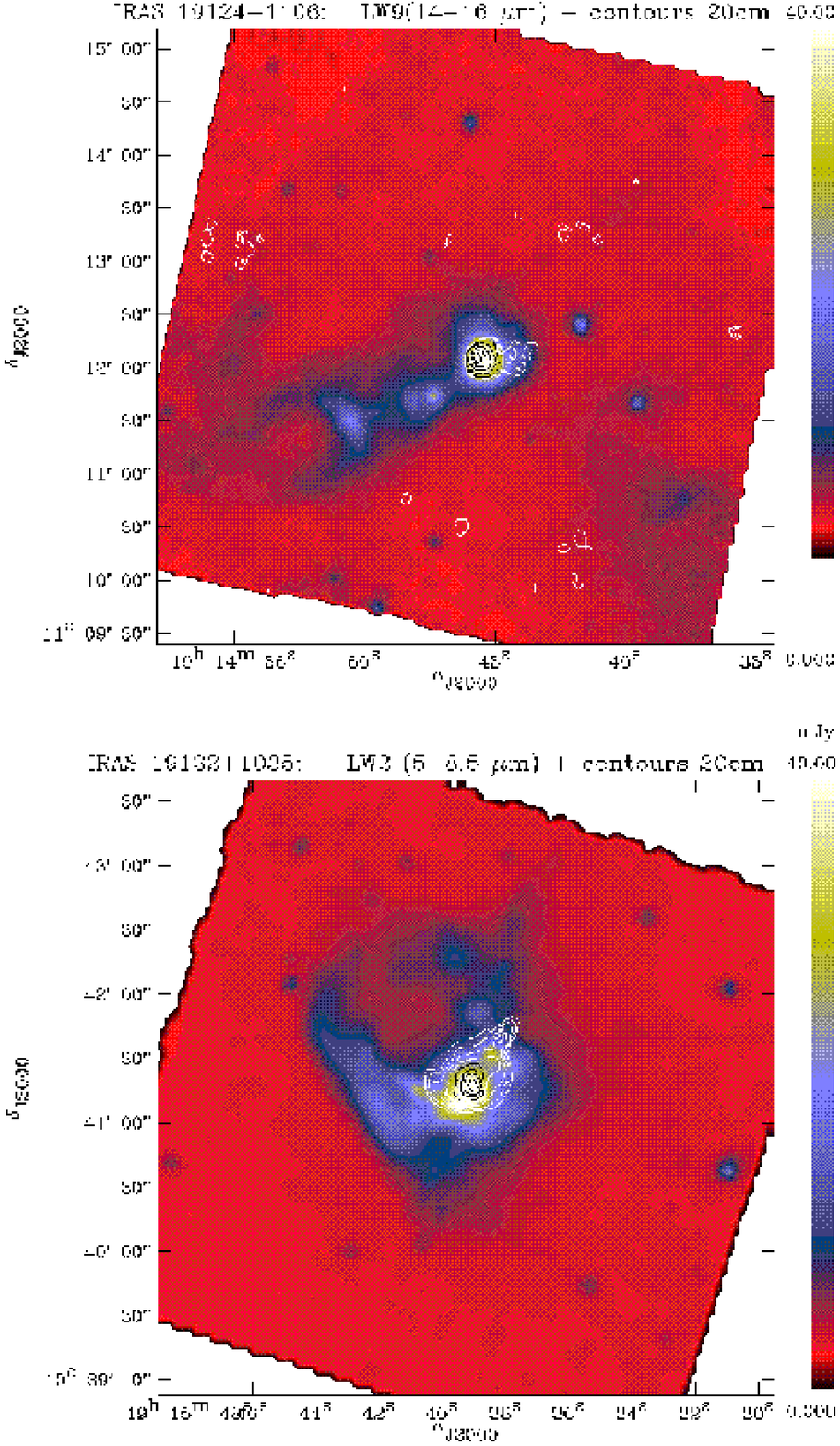}
\end{picture}
\vspace*{0.5cm}
\caption[]{
Top: ISO map of the source $\irasnord$, 
at $\lambda = 15 \microns$,
taken with the LW9 filter. Superimposed are the 20cm radio contours
at 0.4, 0.6, 1, 2, 5, 10 and 15 mJy.
Bottom: ISO map of the source $\irassud$, at $\lambda = 7 \microns$,
taken with the LW2 filter. Superimposed are the 20cm radio contours
at the levels 0.2, 0.4, 0.7, 1.2, 1.8, 2.5, 3, 4 and 5 mJy.
}\label{lobes_iso} 
\end{figure}

\begin{figure}
\setlength{\unitlength}{1.0cm}
\begin{picture}(8,20.)(0.5,1.)
\includegraphics{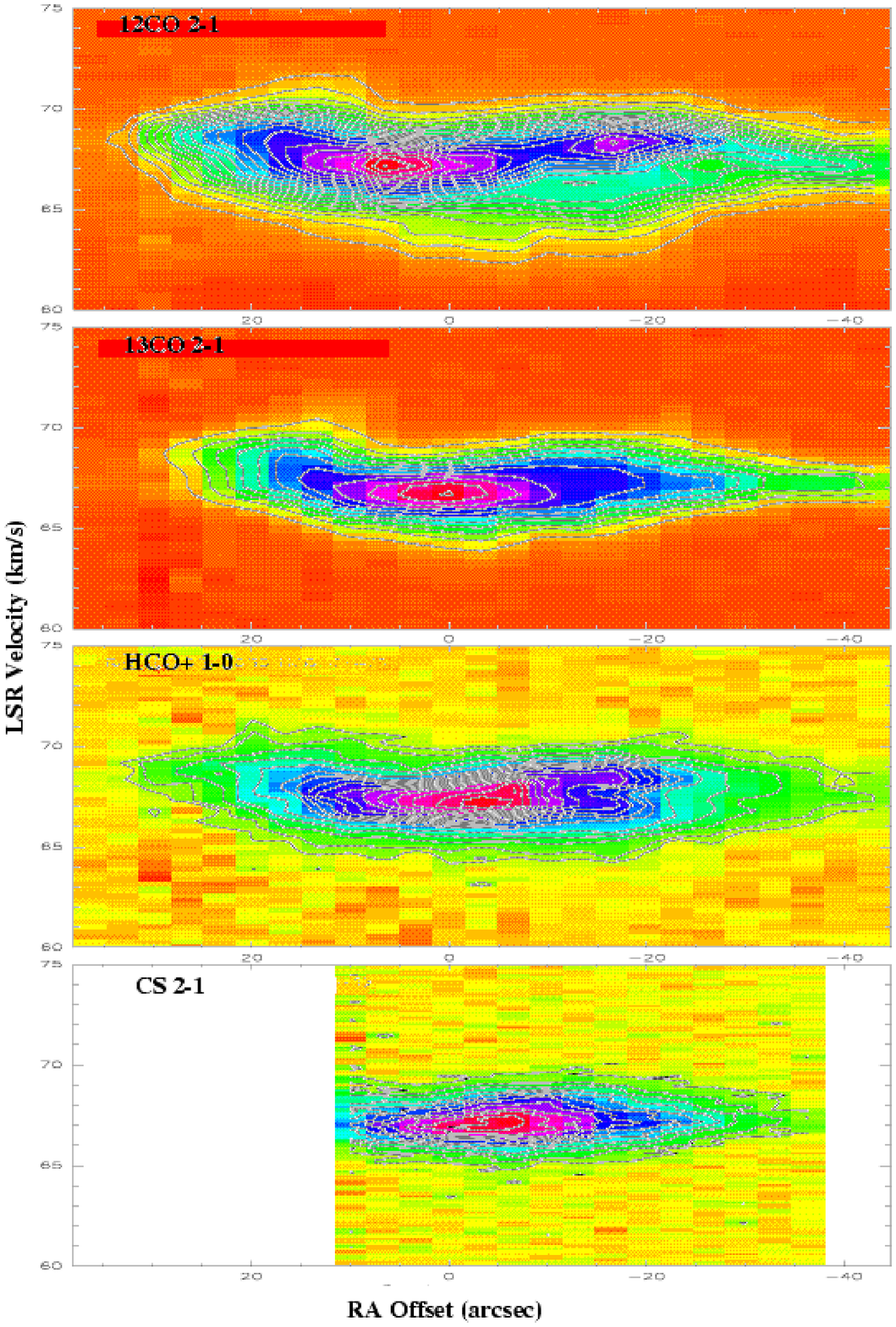}
\end{picture}
\caption[]{
Observations of IRAS 19132+1035.
Offsets are relative to the position of maximum radio emission
observed at the VLA.
The black contours are antennae iso-temperature: 
$\douzecoabb$: $\Ta = -1$, and from $1$ to $20 K$ 
separated by an interval of $1 K$\-; 
$\treizecoabb$: $\Ta = -1$, and from $1$ to 11 K 
separated by an interval of 1 K\-;
$\hcoabb$: $\Ta = -1$, and from $0.2$ to $2.1 K$ 
separated by an interval of $0.1 K$\-; 
$\csabb$: $\Ta = -1$, and from $0.2$ to $2.2 K$ 
separated by an interval of $0.2 K$.
}\label{iras_se_strip2}
\end{figure}

\begin{figure}
\setcounter{figure}{9}
\setlength{\unitlength}{1.0cm}
\begin{picture}(8,20)(1.,1.)
\includegraphics{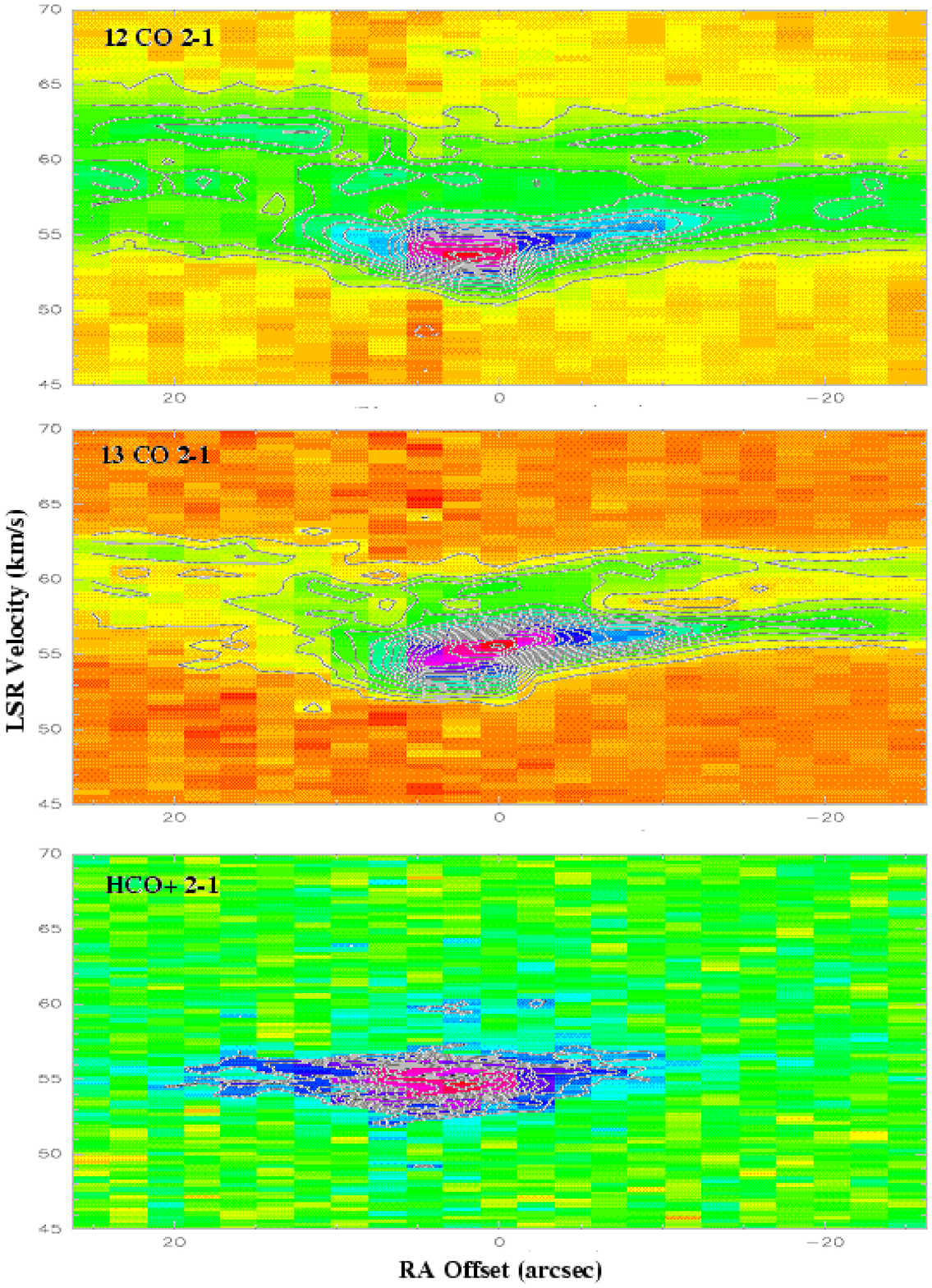}
\end{picture}
\caption[]{
Observations of IRAS 19124+1106. 
Offsets are relative to the position of maximum radio emission
observed at the VLA.
The black contours are antennae iso-temperatures:
$\douzecoabb$: 
$\Ta$ = -1, and from 1 to 14 K separated by an interval of 1 K\-; 
$\treizecoabb$: 
-1 and from 0.5 to 8 K separated by an interval of 0.5 K\-; 
$\hcoabb$: 
-1 0.3 0.4 and from 0.5 to 0.8 K 
separated by an interval of 0.05 K.}\label{iras_nw_strip1}
\end{figure}

\begin{figure}[BH]
\setcounter{figure}{11}
\setlength{\unitlength}{1.0cm}
\begin{picture}(8,20)(0.5,1.)
\includegraphics{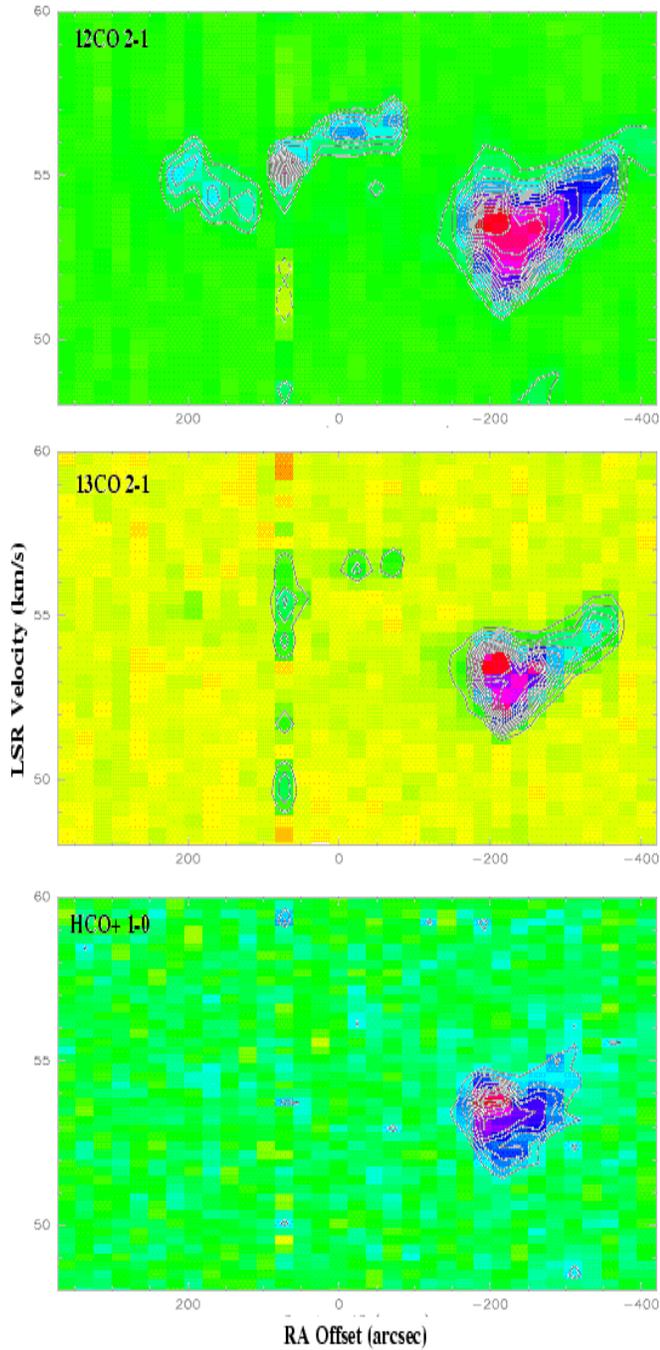}
\end{picture}
\caption[]{
Observations of W50, the supernova remnant shell surrounding SS 433.
Offsets are relative to the position of the central source.  
The black contours are antennae iso-temperatures:
$\douzecoabb$: 
$\Ta$ = -1 and from 1 to 12 $\K$ separated by 1 K\-; 
$\treizecoabb$: 
-1 and from 0.5 to 5 K separated by 1 K\-; 
$\hcoabb$: 
-1 and from 0.2 to 1 K separated by 0.1 K.
}\label{w50_strip}
\end{figure}

\end{document}